\documentclass[
  reprint,            
  superscriptaddress,
  aps,
  pra,
  amsmath,amssymb,
  longbibliography
]{revtex4-2}

\usepackage{graphicx}   
\usepackage{dcolumn}    
\usepackage{bm}         
\usepackage{xcolor}
\usepackage{siunitx}    

\usepackage[colorlinks=true,allcolors=blue]{hyperref}

\begin{document}

\title{Magnet-Free Nonreciprocal frequency conversion using Sequential Temporal modulation: Theory and Simulations}

\author{Arya G.~Pour}
\email{aryag@vt.edu}
\affiliation{Physics Department and Center for Quantum Information Science and Engineering (VTQ), Virginia Tech, Blacksburg, VA 24061, USA}

\author{Jun Ji}
\email{junji@vt.edu}
\affiliation{Bradley Department of Electrical and Computer Engineering, Virginia Tech, Blacksburg, VA 24061, USA}

\author{Linbo Shao}
\email{shaolb@vt.edu}
\affiliation{Physics Department and Center for Quantum Information Science and Engineering (VTQ), Virginia Tech, Blacksburg, VA 24061, USA}
\affiliation{Bradley Department of Electrical and Computer Engineering, Virginia Tech, Blacksburg, VA 24061, USA}

\date{\today}

\begin{abstract}
Nonreciprocal conversion is essential for protecting sources and enabling unidirectional signal routing in photonic, phononic, electronics, and quantum systems, yet conventional implementations rely on magnetic bias that could be challenging to integrate on chip. We propose a magnet-free scheme for frequency-domain nonreciprocity based on sequential, time-gated couplings in a three-mode system. By activating interactions in a fixed temporal order, the forward and reverse frequency conversion pathways acquire unequal dwell times in a lossy intermediate mode, producing strong nonreciprocity without requiring nonlinearities or magnetic materials. Using a harmonic-balance formulation and a Dyson-Born expansion, we derive a compact analytical expression for the isolation ratio that reveals the roles of Floquet sidebands, duty-cycle control, modulation frequency, and dissipation. The results are confirmed by direct time-domain simulations over a wide parameter range. From these results, we extract practical design rules for optimizing isolation through temporal sequencing, loss engineering, and modulation timing. The framework is general and directly applicable to integrated platforms in photonics, phononics, microwave electronics, and superconducting circuits.
\end{abstract}

\maketitle

\section{Introduction}

Nonreciprocal devices are central to photonic, phononic, electronics, and quantum platforms \cite{PT_limits,ClerkPRX}. Isolators suppress back-reflections, protecting fragile sources such as superconducting qubits from feedback-induced instabilities, while circulators allow for sending and receiving signals at the same time on a single channel \cite{MicrowaveCirc,ClerkPRX}. 
In time-invariant systems, Lorentz reciprocity ensures a symmetric scattering matrix; achieving isolation therefore requires breaking time-reversal symmetry \cite{Haldane2008PRL,Onsager1931}. 
While magneto-optic gyrotropy provides robust isolation, it relies on bulky magnets and specialized materials that hinder Complementary Metal-Oxide-Semiconductor (CMOS) compatibility \cite{Faraday,Stadler2014IEEEPhotJ}, motivating a broad search for magnet-free strategies.

To overcome these integration challenges, a wide range of magnet-free strategies have been explored, though often with significant trade-offs. Approaches leveraging optical nonlinearities \cite{Fan_Kerr_diode,Chi2_iso} are compact but susceptible to dynamic reciprocity and noise \cite{DynamicReciprocity}. 
Optomechanical and Brillouin systems achieve nonreciprocity via phonon-mediated scattering \cite{HafeziRabl,Dong2015NatPhoton_BSIT,Tian2021NP}, offering quantum compatibility but limited utility for frequency-domain processing tasks \cite{OMIT_limits}. 
Similarly, topological photonics \cite{Wang2009,Rechtsman2013}, reservoir engineering \cite{Bernier2017NatCommun_OMCirc}, and motion-biased media \cite{Sagnac_nonrecip} provide fundamental asymmetry but often face hurdles regarding integration complexity\cite{STM_limits}.

Spatiotemporal modulation has emerged as a promising magnet-free 
alternative, using traveling-wave refractive index modulations that 
simultaneously impart frequency and wavevector shifts to drive 
indirect interband transitions \cite{YuFan09,Lira12,STM_limits}. 
These implementations rely on strict velocity-matching and impedance 
constraints \cite{Doerr2014OE_SiBroadband,Wang2017IPTL}. 
Parametrically modulated resonators and Floquet-engineered few-mode 
systems have also demonstrated spatial isolation and circulation 
through symmetry-breaking mechanisms 
\cite{Ranzani2015Graph,Kamal2017Parametric,Bernier2017NatCommun_OMCirc}. 
However, all of these approaches predominantly target spatial routing 
between ports at the same frequency. Frequency-domain nonreciprocity 
remains comparatively underexplored, and existing demonstrations of 
nonreciprocal frequency conversion rely on gauge-phase interference 
in multi-path optomechanical plaquettes 
\cite{PhysRevLett.130.013601}, requiring simultaneous balancing of 
multiple control tones and precise phase tuning across an 
optomechanical network. This capability is critical for quantum 
information architectures, where controlled frequency conversion 
between non-degenerate modes enables the routing, multiplexing, 
and isolation required for scalable quantum systems 
\cite{Ranzani2015Graph,Kamal2017Parametric}.

In this work, we propose a magnet-free scheme for frequency-domain 
nonreciprocity based on sequential, time-gated couplings in a 
three-mode system, where the modes may represent optical resonances, 
phononic eigenmodes, microwave cavity fields, or spin sublevels.
Directionality arises from the fixed temporal ordering of discrete 
gate pulses rather than from momentum matching, removing 
velocity-matching constraints entirely and making the scheme 
compatible with CMOS-integrated photonic, phononic, and 
superconducting platforms.
We derive an analytical framework, validated by time-domain 
simulations, that reveals how Floquet sidebands, duty cycles, and 
dissipation govern performance and provides practical design 
guidelines for optimizing frequency-based isolation.

\begin{figure}[t]
  \centering

  \begin{minipage}[b]{\linewidth}
    \centering
    \includegraphics[width=\linewidth]{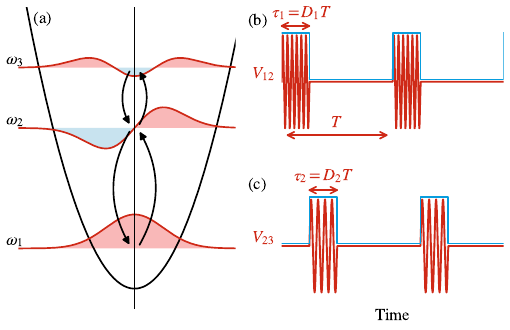}

    \label{fig:schem}
  \end{minipage}

  \caption{%
(a) Energy-level schematic of nonreciprocal frequency conversion. 
The forward operation: a pump at $\omega_{1}$ drives output at $\omega_{3}$ via 
$1\!\to\!2\!\to\!3$; the reverse operation: a pump at $\omega_{3}$ produces output 
at $\omega_{1}$ via $3\!\to\!2\!\to\!1$. (b) Time-domain gate waveform $g_{12}(t)$ 
(blue), active for $0\!\le\! t\!<\!\tau_{1}$; the red trace indicates the gating 
signal $V_{12}(t)$. (c) Time-domain gate waveform $g_{23}(t)$ (blue), active for 
$\tau_{1}\!\le\! t\!<\!\tau_{1}+\tau_{2}$; the red trace indicates the gating 
signal $V_{23}(t)$.}
  \label{fig:schem-two-panels}
\end{figure}

\section{Theory}
\label{sec:theory}
We study a three-mode system with state $\bm a(t)=[a_1,a_2,a_3]^\top$. Its dynamics obey
\begin{equation}
i\,\dot{\bm a}(t)=H(t)\,\bm a(t)+iF\,e^{-i\omega_p t}\,\bm u,
\end{equation}
where $F$ is the driving amplitude at frequency $\omega_p$ and $\bm u$ specifies how the drive couples to the modes.
We split the Hamiltonian into a stationary diagonal part $H_0$ and a time-dependent interaction $V(t)$,

\begin{align}
H(t) &= H_0 + V(t), \\
H_0  &= \operatorname{diag}\!\big(\omega_1 - i\kappa_1/2,\ 
         \omega_2 - i\kappa_2/2,\ \omega_3 - i\kappa_3/2\big), \\
V(t) &= -\,g_{12}(t)\cos(\Omega_1 t)\,E_{12}
        -\,g_{23}(t)\cos(\Omega_2 t)\,E_{23}, \\
E_{12} &= |1\rangle\langle2|+|2\rangle\langle1|, \quad
E_{23}  = |2\rangle\langle3|+|3\rangle\langle2|,
\end{align}

where $\omega_i$ is the resonance frequency of mode $i$, 
$\kappa_i$ its decay rate, $g_{12}(t)$ and $g_{23}(t)$ are 
time-dependent coupling amplitudes, and $\Omega_1$, $\Omega_2$ 
are the respective modulation frequencies.
The coupling operators $E_{12}$ and $E_{23}$ restrict interactions 
to the $1\text{--}2$ and $2\text{--}3$ pairs, as illustrated in 
Fig.~\ref{fig:schem-two-panels}.
The couplings are activated sequentially: $g_{12}(t)$ is on during 
$[0,\tau_1)$, $g_{23}(t)$ during $[\tau_1, \tau_1+\tau_2)$, and 
both are off in the remainder $[\tau_1+\tau_2, T)$, where $T=2\pi/\Omega$ 
is the modulation period.
In terms of the normalized duty cycles $D_1 = \tau_1/T$ and 
$D_2 = \tau_2/T$, the gate profiles are

\begin{equation}
g_{12}(t) = g_{12[0,\tau_{1})}(t), \qquad
g_{23}(t) = g_{23[\tau_{1},\tau_{1}+\tau_{2})}(t \bmod T).
\label{eq:gating}
\end{equation}

\paragraph{Time-reversal symmetry breaking.}
A Hamiltonian breaks time-reversal symmetry if and only if $H^*(-t) \neq H(t)$. 
For a static Hamiltonian this reduces to $H_0^* = H_0$. In magneto-optic systems, 
a static magnetic bias introduces imaginary antisymmetric off-diagonal gyrotropy 
terms; since complex conjugation reverses the sign of these terms, $H_0^* \neq H_0$ 
and time-reversal symmetry is broken without any time 
dependence~~\cite{Faraday,Stadler2014IEEEPhotJ,Haldane2008PRL,Onsager1931,Onsager1931b,Ballantine1929,Carson1929}. Here, by contrast, $H_0$ is real and symmetric, and 
nonreciprocity is produced entirely by $V(t)$: activating $g_{12}(t)$ and $g_{23}(t)$ 
in a fixed temporal order within each period makes $V^*(-t) \neq V(t)$, and therefore 
$H^*(-t) \neq H(t)$. This ordering establishes a preferred direction: the forward 
conversion $\omega_1\!\to\!\omega_3$ proceeds through $g_{12}$ followed by $g_{23}$, 
whereas the reverse process $\omega_3\!\to\!\omega_1$ encounters the opposite ordering 
and is suppressed, making the temporal sequencing the sole origin of nonreciprocity 
in our scheme.

\begin{figure}[t]
  \centering
  \includegraphics[width=\linewidth]{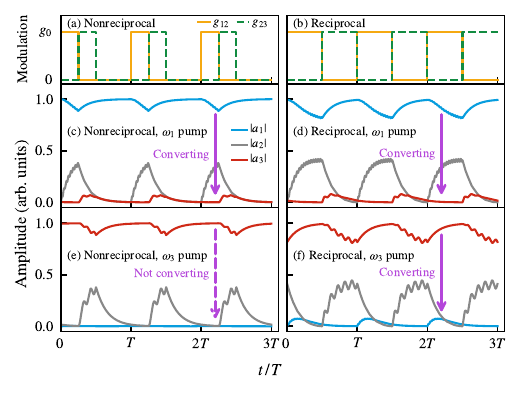}
  \caption{Time-domain modes $|a_j(t)|$ in the steady-state regime. Panels (a) and (b) show the gate modulations $g_{12}(t)$ and 
$g_{23}(t)$ for the nonreciprocal sequential protocol 
($D_1 = D_2 = 0.25$) and the reciprocal reference 
($\tau_1 = \tau_2 = T/2$), respectively. 
Panels (c) and (e) show the mode amplitudes under nonreciprocal 
modulation with pump at $\omega_1$ and $\omega_3$ respectively: 
the forward process (c) produces clear conversion from $\omega_1$ 
to $\omega_3$, while the reverse process (e) shows no conversion, 
confirming nonreciprocal behavior. 
Panels (d) and (f) show the reciprocal reference under the same 
pump directions: both forward and reverse processes yield 
conversion, confirming equal forward and reverse conversion.}
  \label{fig:recip_vs_nonrecip}
\end{figure}

\paragraph{Reciprocal versus nonreciprocal response.}
Before developing the formal analysis, we illustrate the key 
contrast between reciprocal and nonreciprocal modulation through 
direct time-domain simulations.
The reciprocal reference case is obtained by setting 
$\tau_1 = \tau_2 = T/2$, so that $g_{12}$ is active during 
$[0, T/2)$ and $g_{23}$ during $[T/2, T)$, together covering 
the full modulation period without any idle interval.
The two gates are then related by a simple time shift of $T/2$, 
making the protocol invariant under time reversal and yielding 
equal forward and reverse conversion amplitudes.
The nonreciprocal case corresponds to the sequential protocol 
of Eq.~\eqref{eq:gating}, where $g_{12}$ is active during 
$[0,\tau_1)$, $g_{23}$ activates during $[\tau_1,\tau_1+\tau_2)$, 
and both couplings are off in the remaining interval 
$[\tau_1+\tau_2, T)$.
Fig.~\ref{fig:recip_vs_nonrecip} shows the gate waveforms and 
steady-state mode amplitudes $|a_j(t)|$ for both the nonreciprocal 
and reciprocal protocols under forward and reverse pumping. 
Panels (a) and (b) illustrate the temporal structure of the two 
modulation schemes. In the nonreciprocal case, panels (c) and (e) 
reveal a clear asymmetry: pumping at $\omega_1$ produces 
significant conversion to $\omega_3$, while pumping at $\omega_3$ 
produces no conversion to $\omega_1$, confirming nonreciprocal 
frequency conversion. In the reciprocal case, panels (d) and (f) 
yield conversion under both pump directions with nearly equal 
amplitudes, confirming that time-reversal symmetry is preserved 
when the two gates together cover the full modulation period 
without an idle interval. The asymmetry in the nonreciprocal case 
originates entirely from the temporal ordering of the gates, not 
from any difference in coupling strengths or mode frequencies, 
and the formal derivation that follows explains this contrast 
through Floquet sideband interference.

\paragraph{Expansion in Fourier space}
Having established that $V(t)$ breaks time-reversal symmetry, 
we now develop the analytical framework to quantify the 
resulting nonreciprocity.

We therefore expand
\begin{equation}
\bm a(t)=\sum_{n\in\mathbb Z}\bm a^{(n)}\,e^{-i(\omega_p+n\Omega)t},
\end{equation}
and denote the component on mode $i$ at sideband $n$ by $a_i^{(n)}$. For a single-mode drive on mode $j$ ($\bm u=\bm e_j$) we write the corresponding transfer channel as
\begin{equation}
a^{(n)}_{i\leftarrow j}:=a_i^{(n)}\Big|_{\bm u=\bm e_j}.
\end{equation}

Also, we will assume
\begin{equation}
\Omega_1 = \omega_2 - \omega_1= p_1 \Omega + \delta_{12},\qquad
\Omega_2 = \omega_3 - \omega_2 = p_2 \Omega + \delta_{23}
\end{equation}

where $p_1$ and $p_2$ are whole numbers.
The isolation ratio $\mathcal{R}_{n,n'}$ quantifies the degree of 
nonreciprocity by comparing the forward conversion amplitude (pump at 
$\omega_1$, output at $\omega_3$) against the reverse (pump at $\omega_3$, 
output at $\omega_1$). Expressed in decibels as 
$I = 20\log_{10}\mathcal{R}$, a value $I=0$\,dB corresponds to a 
fully reciprocal system while larger $|I|$ indicates stronger isolation.
\begin{equation}
\mathcal R_{n,n'}:=|\frac{a^{(n)}_{3\leftarrow 1}}{a^{(n')}_{1\leftarrow 3}}|
\label{eq:R_def}
\end{equation}

\paragraph{General construction of $\bm a^{(n)}$ from Green functions and $V^{(m)}$.}

We define the Green function and the Fourier coefficients of the interaction,
\begin{equation}
    \bm G_0(\omega) := [\omega \bm I - H_0]^{-1}, \quad V(t) = \sum_{m \in \mathbb{Z}} V^{(m)} e^{-im\Omega t}.
    \label{eq:G0_def}
\end{equation}

Since the bare Hamiltonian $H_0$ is diagonal, the Green function decomposes into scalar propagators for each mode $j \in \{1,2,3\}$:
\begin{equation}
    G_j(\omega) = \frac{1}{\omega - \omega_j + i\kappa_j/2}.
    \label{eq:Gj_def}
\end{equation}

Substituting the harmonic expansion into the equation of motion yields the recursion
\begin{equation}
    \bm a^{(n)}
    = \bm G_0(\omega_p{+}n\Omega)\!
    \Big(\sum_{m\in\mathbb Z} V^{(m)}\,\bm a^{(n-m)} + F\,\delta_{n0}\,\bm u\Big).
    \label{eq:HB_recursion}
\end{equation}

\emph{Dyson/Born series} \cite{Fetter1971}. In the Dyson expansion, a $1\!\to\!3$ conversion requires two successive interactions, 
since $E_{12}$ couples only $1\leftrightarrow 2$ and $E_{23}$ couples only $2\leftrightarrow 3$. 
The lowest-order nonvanishing contribution is therefore second order, corresponding to two 
frequency hops through the intermediate mode. Explicitly,
\begin{widetext}
\begin{equation}
 \bm a^{(n)}_{(2)}=\bm G_0(\omega_p{+}n\Omega)\Big(\sum_{m} V^{(m)}\,\bm G_0(\omega_p{+}(n{-}m)\Omega)\,V^{(n-m)}\Big)\bm G_0(\omega_p)\,F\bm u .
\end{equation}
\end{widetext}
where $V^{(m)}$ are the Fourier coefficients of the interaction operator.

\paragraph{Fourier spectra of the sequential gates.}
The sequential protocol generates the sideband spectra
\begin{widetext}
\begin{equation}
\begin{aligned}
V_{12}^{(m)}
&=\frac{g_{12}}{2}\!\left[
\frac{e^{\,i\pi(m-p_1)D_{1}}\,\sin\!\big(\pi(m-p_1)D_{1}\big)}{\pi(m-p_1)}
+\frac{e^{\,i\pi(m+p_1)D_{1}}\,\sin\!\big(\pi(m+p_1)D_{1}\big)}{\pi(m+p_1)}
\right],\\[6pt]
V_{23}^{(m)}
&=\frac{g_{23}}{2}\!\left[
\frac{e^{\,i2\pi(m-p_2)D_{1}}\,e^{\,i\pi(m-p_2)D_{2}}\,\sin\!\big(\pi(m-p_2)D_{2}\big)}{\pi(m-p_2)}
+\frac{e^{\,i2\pi(m+p_2)D_{1}}\,e^{\,i\pi(m+p_2)D_{2}}\,\sin\!\big(\pi(m+p_2)D_{2}\big)}{\pi(m+p_2)}
\right].
\end{aligned}
\label{eq:gq_tau}
\end{equation}
\end{widetext}
These expressions encode both the weight and phase of each harmonic.

Using the Green-function recursion, the isolation ratio between the two-hop processes can be written as
\begin{widetext}
\begin{equation}
\mathcal R_{p_1-p_2,\,p_2-p_1}\ \approx\
\frac{G_3(\omega_3 - \delta_{13})\,G_1(\omega_1)}{G_1(\omega_1 + \delta_{13})\,G_3(\omega_3)}\;
\frac{\sum_{m} V_{23}^{(m)}\,V_{12}^{(p_1 - p_2 - m)}\,G_2(\omega_3 + \delta_{13} - m \Omega)}
     {\sum_{m} V_{12}^{(m)}\,V_{23}^{(p_2 - p_1 - m)}\,G_2(\omega_1 - \delta_{13} - m \Omega)}.
\label{eq:General_R}
\end{equation}
\end{widetext}

where $\delta_{13}=\delta_{12}-\delta_{23}$. Under condition $\delta_{13}=0$, this reduces to
\begin{equation}
    \mathcal R \approx\
\frac{\sum_{m} V_{23}^{(m)}\,V_{12}^{(p_1 - p_2 -m)}(\omega_1 - \omega_2 - m \Omega + i \kappa_2 / 2)}{ \sum_{m} V_{12}^{(m)}\,V_{23}^{(p_2 - p_1 - m)}(\omega_3 - \omega_2 - m \Omega + i \kappa_2 / 2)},
\label{eq:Final_R}
\end{equation}

Equation~\eqref{eq:Final_R} expresses the ratio of forward 
to reverse transfer amplitudes under single-mode excitation. 
The overall coupling constants $g_{12}$ and $g_{23}$ cancel 
between numerator and denominator when attention is restricted 
to the sequential $1\!\to\!2\!\to\!3$ and $3\!\to\!2\!\to\!1$ 
pathways, so absolute coupling strengths do not govern the 
isolation at this order.
This cancellation is approximate: higher-order and 
non-sequential contributions reintroduce a residual 
dependence on $g_{12}$ and $g_{23}$, which we neglect 
here for clarity.
Under the condition $\delta_{13} = 0$, the outer 
mode losses $\kappa_1$ and $\kappa_3$ drop out of 
Eq.~\eqref{eq:Final_R}, leaving the intermediate loss 
$\kappa_2$ as the sole dissipative parameter controlling 
isolation.

Having established the physical origin of the isolation, 
we now analyze the sideband structure of Eq.~\eqref{eq:Final_R} 
to identify the dominant contributions.
\paragraph{Dominant terms.}

\begin{figure}
    \includegraphics[width=\linewidth]{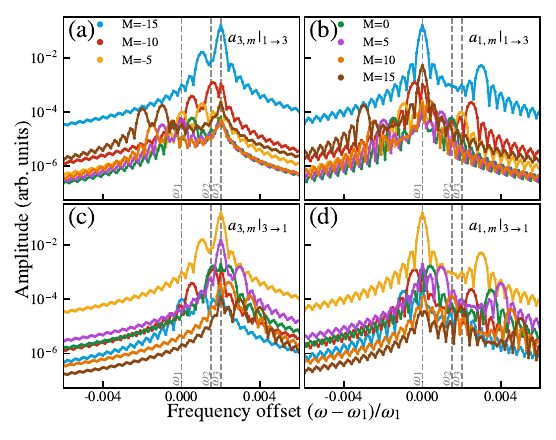}
\caption{Single–$m$ Fourier contributions to the sequential two-hop process.  
(a) $|a_3^{(m)}|$ with pump at $\omega_{1}$, dominated by the $m=-p_{1}$ sideband.  
(b) $|a_1^{(m)}|$ with pump at $\omega_{1}$, also dominated by $m=-p_{1}$.  
(c) $|a_3^{(m)}|$ with pump at $\omega_{3}$, where the leading term comes from $m=-p_{2}$.  
(d) $|a_1^{(m)}|$ with pump at $\omega_{3}$, likewise dominated by $m=-p_{2}$. Mode frequencies were set to 
$\omega_1 = 1.0000$, $\omega_2 = 1.0015$, and $\omega_3 = 1.0020$, 
with quality factors $Q_1 = Q_2 = Q_3 = 5000$, corresponding to decay rates 
$\kappa_i = \omega_i / Q_i$. The drive amplitude was 
$F = 1\times10^{-5}$, coupling amplitudes $g_{12} = g_{23} = 1\times10^{-4}$, 
and the modulation frequency was $\Omega = 1\times10^{-4}$.}
   
    \label{fig:SingleM}
\end{figure}

We now isolate the leading contributions in the two-hop processes in order to identify the dominant terms that can serve as the starting point for expansions. With $n=p_1+p_2$ (so that $\omega_1+n\Omega\approx\omega_3$), the forward amplitude
\[
\sum_m V_{23}^{(m)}\,G_2\!\big(\omega_1{+}(n{-}m)\Omega\big)\,V_{12}^{(n-m)}
\]
is dominated by the leading sidebands, i.e., when the sideband indices match the frequency gaps between modes. This occurs at
 
\[
m^\star=-p_2,\qquad n-m^\star=+p_1,
\]
which places the virtual intermediate frequency at $\omega_2$. The reverse process at $n'=p_2-p_1$ is dominated by the analogous choice

\[
m'{}^\star=-p_1,\qquad n'-m'{}^\star=+p_2.
\]
These leading sidebands coincide with the peaks visible in the single-$m$ decomposition of Fig.~\ref{fig:SingleM}.

At these indices the sideband weights are
\begin{equation}
\begin{aligned}
V_{12}^{(\pm p_1)}
&=\frac{g_{12}}{2}\!\left(
  D_{1}
  + \frac{\sin(2\pi p_{1} D_{1})}{2\pi p_{1}}\,e^{\pm i\,2\pi p_{1} D_{1}}
\right), \\[6pt]
V_{23}^{(\pm p_2)}
&=\frac{g_{23}}{2}\!\left(
  D_{2}
  + \frac{\sin(2\pi p_{2} D_{2})}{2\pi p_{2}}\,e^{\pm i\,2\pi p_{2}(D_{1}+D_{2})}
\right).
\end{aligned}
\end{equation}

Evaluated at exact output alignment 
\((\omega_1+n\Omega=\omega_3,\;\omega_3+n'\Omega=\omega_1)\),
the forward–reverse ratio simplifies to
\begin{widetext}
\[
\mathcal R_{p_1-p_2,\,p_2-p_1}\approx
\frac{V_{23}^{(-p_2)}\,V_{12}^{(+p_1)}}{V_{12}^{(-p_1)}\,V_{23}^{(+p_2)}} =
\frac{\Big[D_{2}+\tfrac{\sin(2\pi p_{2}D_{2})}{2\pi p_{2}}\,e^{-i\,2\pi p_{2}(D_{1}+D_{2})}\Big]
      \Big[D_{1}+\tfrac{\sin(2\pi p_{1}D_{1})}{2\pi p_{1}}\,e^{+i\,2\pi p_{1}D_{1}}\Big]}
     {\Big[D_{1}+\tfrac{\sin(2\pi p_{1}D_{1})}{2\pi p_{1}}\,e^{-i\,2\pi p_{1}D_{1}}\Big]
      \Big[D_{2}+\tfrac{\sin(2\pi p_{2}D_{2})}{2\pi p_{2}}\,e^{+i\,2\pi p_{2}(D_{1}+D_{2})}\Big]}.
\]
\end{widetext}
Taking magnitudes yields
\[
\mathcal R_{p_1-p_2,\,p_2-p_1}=1,
\]
As a result, keeping only the leading sideband contributions yields no isolation. Directionality instead emerges from neighboring sidebands, whose interference breaks the forward–reverse symmetry. Expanding about the dominant sidebands provides a systematic way to capture these asymmetry-inducing corrections.
\paragraph{Single–$m$ Fourier decomposition.}

Fig.~\ref{fig:SingleM} shows the single–$m$ Fourier components contributing to the two-hop process under forward and reverse excitation.
Panels~(a) and~(b) correspond to forward drive with the pump applied at $\omega_{1}$.
In (a), the response of mode~3, $|a_{3}|$, peaks near $\omega_{3}$, indicating that energy transfer proceeds through the $m=-p_{1}$ sideband.
Similarly, in (b), the amplitude of mode~1, $|a_{1}|$, is dominated by the same $m=-p_{1}$ component.
Panels~(c) and~(d) show the reverse process with the pump applied at $\omega_{3}$.
Here, $|a_{3}|$ in (c) and $|a_{1}|$ in (d) both exhibit dominant contributions around $m=-p_{2}$, corresponding to transfer through the opposite sideband.
Summing over all $m$ yields the net steady-state amplitudes: the total spectral weight of mode~1 in (d) exceeds that of mode~3 in (a), revealing the asymmetry responsible for nonreciprocal energy transfer.
This imbalance originates from unequal constructive and destructive interference among adjacent sidebands in the forward and reverse pathways.

The net steady-state amplitudes extracted from this 
sideband analysis are compared directly against 
time-domain simulations in Sec.~\ref{sec:results}, 
confirming the accuracy of the perturbative framework.
\section{Results}
\label{sec:results}

In Sec.~\ref{sec:theory} we derived an analytical expression for the energy transfer between modes that determines the isolation ratio [Eq.~\eqref{eq:Final_R}] using a Fourier-Floquet expansion of a sequentially modulated three-mode system. We now evaluate the accuracy of this expression by comparing it with direct time-domain simulations for an experimentally relevant observable that combines the response from all modes.
For the forward configuration, the drive is applied to mode~1 and the summed response across modes is measured at the frequency of mode~3. For the reverse configuration, the drive is applied to mode~3 and measured at mode~1.
We simulate two forms of the time-dependent Hamiltonian. The first is the effective sequential-gating model derived in Sec.~\ref{sec:theory}, where only one coupling is active during each modulation window: the $1{\leftrightarrow}2$ coupling in the first gate and the $2{\leftrightarrow}3$ coupling in the second. The second is a resonant model that more closely represents experimental conditions, where a residual overlap between couplings is present within each active gate, allowing both $1{\leftrightarrow}2$ and $2{\leftrightarrow}3$ interactions to occur simultaneously.

The interaction used in resonant model is expressed as
\[
\mathbf{V}(t) =
\begin{cases}
g_{12}\cos(\Omega_1 t)
\begin{bmatrix}
0 & 1 & 0\\
1 & 0 & 1\\
0 & 1 & 0
\end{bmatrix}, & 0 < t < \tau_1,\\[8pt]
g_{23}\cos(\Omega_2 t)
\begin{bmatrix}
0 & 1 & 0\\
1 & 0 & 1\\
0 & 1 & 0
\end{bmatrix}, & \tau_1 < t < \tau_1 + \tau_2.
\end{cases}
\]

For analytical calculations, we use the Fourier spectra of the gated couplings. The corresponding components are
\[
V_{12}^{(m)} = \frac{g_{12}}{2}\Big[\,W(m - p_1; D_1, 0) + W(m + p_1; D_1, 0)\,\Big],
\]
\[
V_{23}^{(m)} = \frac{g_{23}}{2}\Big[\,W(m - p_2; D_2, D_1) + W(m + p_2; D_2, D_1)\,\Big],
\]
where
\[
W(m; w, s) =
\begin{cases}
w, & m = 0,\\[6pt]
e^{i2\pi m (s + \tfrac{w}{2})}\,\dfrac{\sin(\pi m w)}{\pi m}, & m \neq 0~.
\end{cases}
\]

Here, $w$ is the fractional width of the gate, $s$ is its fractional start time, and $p_1, p_2$ are the carrier indices associated with the two modulation frequencies.
These Fourier components are substituted into the harmonic-balance recursion [Eq.~(\ref{eq:HB_recursion}); see Appendix~\ref{Apx:HB}] to obtain the analytical isolation, which we compare directly to the time-domain simulation results for both the effective and resonant models.

\subsection{Comparison with Time-domain Simulations}

\begin{figure}
    \centering
    \includegraphics[width=\linewidth]{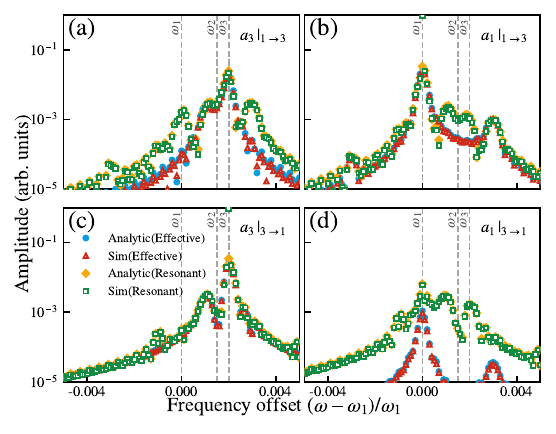}
\caption{Floquet-line amplitudes from the analytical two-hop model (dots) and time-domain simulation (triangles).  
(a) $|a_{3}|$, pump at $\omega_{1}$, showing the main response at $\omega_{3}$.  
(b) $|a_{1}|$, pump at $\omega_{1}$, with the main response at $\omega_{1}$.  
(c) $|a_{3}|$, pump at $\omega_{3}$, with the main response at $\omega_{1}$.  
(d) $|a_{1}|$, pump at $\omega_{3}$, showing the main response at $\omega_{3}$.  
Simulation parameters: Mode frequencies were set to 
$\omega_1 = 1.0000$, $\omega_2 = 1.0015$, and $\omega_3 = 1.0020$, 
with quality factors $Q_1 = Q_2 = Q_3 = 5000$, corresponding to decay rates 
$\kappa_i = \omega_i / Q_i$. The drive amplitude was 
$F = 1\times10^{-5}$, coupling amplitudes $g_{12} = g_{23} = 1\times10^{-4}$, 
and the modulation frequency was $\Omega = 1\times10^{-4}$.}
    \label{fig:sim}
\end{figure}

To validate our theoretical framework, we numerically integrated the time-domain equations of motion for the driven system with sequential couplings $g_{12}(t)$ and $g_{23}(t)$ (see Appendix~\ref{Apx:sim}).
The dynamics were integrated in a rotating frame, and the steady-state Floquet sideband amplitudes were extracted by Fourier transforming the system response over a single modulation period. Fig.~\ref{fig:sim} presents the isolation ratio predicted by Eq.~\ref{eq:Final_R} alongside these time-domain simulation results.

The analytical model demonstrates quantitative agreement with the numerical data across the examined ranges of duty cycles ($D_1, D_2$), modulation frequency ($\Omega$), and intermediate mode loss ($\kappa_{2}$).
This correspondence confirms that the perturbative expansion accurately describes the nonreciprocal transport mechanism within this parameter regime.

\subsection{Parameter Dependence of Isolation}

Unless stated otherwise, we fix $(\omega_1,\omega_2,\omega_3)=(1.0000,\,1.0015,\,1.0020)$, set $\kappa_i=\omega_i/Q$, where  quality factor $Q=5000$ and choose couplings $g_{12}=g_{23}=g = 10^{-4}$.  
The duty cycle is taken as $D_1 = D_2 =0.25$, with $(p_1,p_2)=(15,5)$ and modulation frequency $\Omega=1\times10^{-4}$, such that $p_1\Omega\simeq\omega_2-\omega_1$ and $p_2\Omega\simeq\omega_3-\omega_2$.  

For isolation we compare single-mode drives: the forward process uses $\bm u=(1,0,0)$ (pump at $\omega_1$), while the reverse process uses $\bm u=(0,0,1)$ (pump at $\omega_3$).  
Although employing a common drive $\bm u=(1,1,1)$ can yield larger apparent isolation, it simultaneously excites mode~2 and opens single-hop channels $2\!\to\!1$ and $2\!\to\!3$.  
Such additional contributions obscure the interpretation of the ratio and are therefore excluded; all reported results use the single-mode drives defined above.

\begin{figure}[t]
  \centering
  \includegraphics[width=\linewidth]{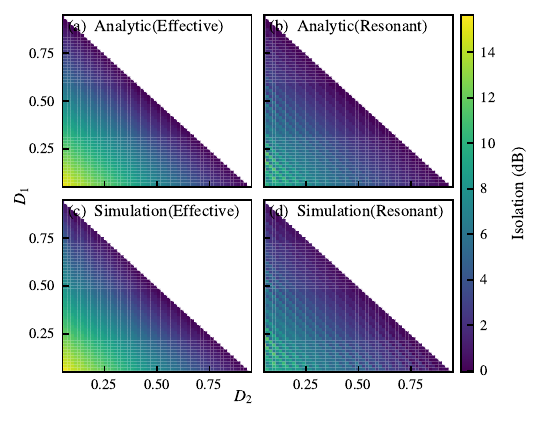}
\caption{
Isolation versus duty cycles $D_1$ and $D_2$. 
(a) Analytic prediction with separate gating windows. 
(b) Resonant analytic prediction captures overlapping interactions. 
(c) Time-domain simulation using the effective Hamiltonian. 
(d) Simulation using the resonant Hamiltonian, which includes simultaneous coupling during switching transitions. 
The resonant model yields lower isolation at small $D$, as both interactions partially overlap in time, reducing the directional contrast.}  \label{fig:iso_tau1_tau2}
\end{figure}

\paragraph{Duty cycle $D$.}
Fig.~\ref{fig:iso_tau1_tau2} shows the isolation as a function 
of the duty cycles. 
Narrowing either window ($D_1$ or $D_2$ small) drives the gates 
toward a delta-like limit with a broad Fourier spectrum. 
This spectrum enables an energy relay along the forward path, 
while the reverse path remains trapped in the lossy intermediate 
mode, suppressing the reverse amplitude and increasing isolation. 
At the boundaries ($D_1=0$, $D_2=0$, $D_1=1$, or $D_2=1$), one 
coupling is absent and two-hop conversion is forbidden, causing 
isolation to vanish. Notably, along the diagonal $D_1 = D_2$ with $D_1 + D_2 = 1$ 
(i.e., $D_1 = D_2 = 0.5$), isolation also vanishes: this 
corresponds exactly to the reciprocal reference case of 
Fig.~\ref{fig:recip_vs_nonrecip}, where the two gates together 
cover the full modulation period without any idle interval, 
restoring time-reversal symmetry and yielding $\mathcal{R}=1$.

\begin{figure}[t]
  \centering
  \includegraphics[width=\linewidth]{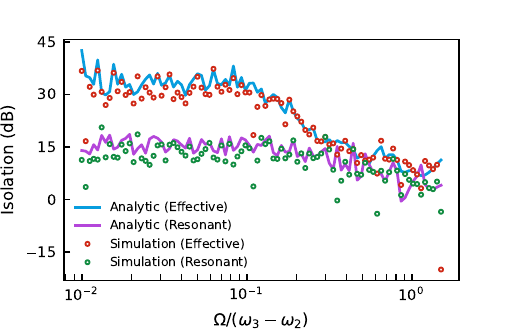}
\caption{
Isolation versus normalized modulation frequency $\Omega/(\omega_{3}-\omega_{2})$. 
The blue line show the analytic prediction with separate gating windows, 
the purple line represent the resonant analytic prediction, 
the red circles correspond to time-domain simulation using the effective Hamiltonian, 
and the green circles indicate simulation with the resonant Hamiltonian. 
At small $\Omega$, isolation increases due to longer dwell times and stronger directional contrast, 
while at large $\Omega$, sidebands move off resonance and isolation decreases, 
highlighting the trade-off between dwell-time suppression and spectral overlap.}
  \label{fig:iso-vs-omega}
\end{figure}

\paragraph{Modulation frequency $\Omega$.}
Fig.~\ref{fig:iso-vs-omega} shows the isolation as a function of the modulation frequency $\Omega$, normalized by $\omega_{3}-\omega_{2}$. 
Consistent with Eq.~(\ref{eq:Final_R}), reducing $\Omega$ (corresponding to a longer modulation period $T$) initially enhances isolation by increasing the asymmetry between forward and reverse frequency-conversion pathways. 
As $\Omega$ decreases, population transferred into the intermediate mode from the reverse direction experiences increased dissipation before it can be converted in the subsequent modulation segment, thereby suppressing the reverse process. 
However, once $\Omega$ becomes small compared to the relaxation rate of the intermediate mode, the dynamics enter an adiabatic regime in which the system relaxes within each gated segment and follows the instantaneous Hamiltonian. 
In this regime, further reducing $\Omega$ does not introduce additional directionality, and the isolation becomes limited by the dissipative bottleneck of the intermediate mode rather than by the modulation rate. 
As a consequence, the isolation saturates and remains approximately constant at sufficiently small $\Omega$.

\begin{figure}[t]
  \centering
  \includegraphics[width=\linewidth]{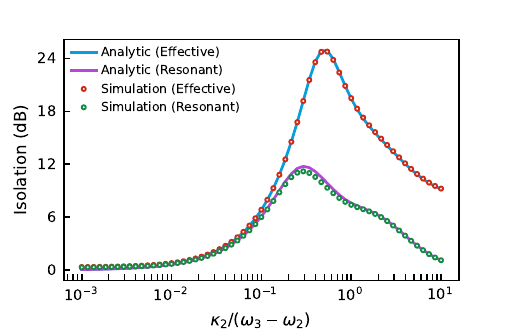}
\caption{
Isolation versus normalized intermediate mode loss $\kappa_2/(\omega_{3}-\omega_{2})$. 
Blue line show the analytic prediction with separate gating windows, 
purple line represent the resonant analytic prediction, 
red circles correspond to simulation using the effective Hamiltonian, 
and green circles indicate simulation with the resonant Hamiltonian. 
Isolation increases with $\kappa_2$ as the reverse path is more strongly damped in the intermediate mode, 
peaks near $\kappa_2\!\sim\!|\omega_{3}-\omega_{2}|$, and decreases for larger losses as all mode amplitudes diminish.}
  \label{fig:iso-vs-kappa2}
\end{figure}

\paragraph{Intermediate mode loss $\kappa_2$ and dwell-time interpretation.}
Fig.~\ref{fig:iso-vs-kappa2} shows the isolation as a function 
of the normalized intermediate mode loss $\kappa_2/(\omega_3-\omega_2)$. 
The dependence on $\kappa_2$ reflects a competition between two 
opposing roles of dissipation in the sequential modulation.
The physical origin of isolation is the asymmetry in the time 
energy spends in mode~2 between forward and reverse processes. 
In the forward direction, $g_{12}$ is immediately followed by 
$g_{23}$, so energy deposited in mode~2 is transferred to 
mode~3 within the gate duration $\tau_1$ before significant 
dissipation occurs. In the reverse direction, energy arrives 
during the $g_{23}$ window and must wait through the full idle 
interval $T - \tau_1 - \tau_2$ before $g_{12}$ activates. 
The dissipation timescale of mode~2 is set directly by the 
pole of the Green function $G_2(\omega) = (\omega - \omega_2 + 
i\kappa_2/2)^{-1}$, which gives $\tau_\mathrm{diss} = 2/\kappa_2$. 
Maximum isolation occurs when $\tau_\mathrm{diss}$ is comparable 
to the idle interval $T - \tau_1 - \tau_2$. For $D_1 = D_2 = 0.25$ 
this interval equals $T/2 = \pi/\Omega$, giving the peak condition
\begin{equation}
    \frac{2}{\kappa_2} \sim \frac{\pi}{\Omega}
    \quad\Longrightarrow\quad
    \frac{\kappa_2}{\omega_3-\omega_2} \sim 
    \frac{2\Omega}{\pi(\omega_3-\omega_2)} 
    = \frac{2}{5\pi} \approx 0.13,
\end{equation}
consistent with the peak observed near 
$\kappa_2/(\omega_3-\omega_2) \approx 0.3$ in 
Fig.~\ref{fig:iso-vs-kappa2}.
In the limit $\kappa_2 \to 0$, the intermediate mode is 
effectively lossless and $\tau_\mathrm{diss} \gg T$. 
Energy deposited in mode~2 persists across many modulation 
periods, so both forward and reverse processes transfer 
energy with equal efficiency. The directional asymmetry 
vanishes and isolation tends to zero.
In the opposite limit $\kappa_2 \to \infty$, $\tau_\mathrm{diss} 
\ll \tau_1$ and mode~2 decays far faster than any gate duration. 
Both forward and reverse processes lose their energy in mode~2 
before any transfer can occur, suppressing mode amplitudes 
equally in both directions and again driving isolation to zero.
The peak isolation therefore occurs at the intermediate value 
of $\kappa_2$ where the reverse path is strongly penalized 
while the forward path retains sufficient amplitude for 
efficient conversion, a balance that is encoded in the 
asymmetric arguments of $G_2(\omega)$ in the numerator 
and denominator of Eq.~\eqref{eq:Final_R}.

\begin{figure}[t]
  \centering
  \includegraphics[width=\linewidth]{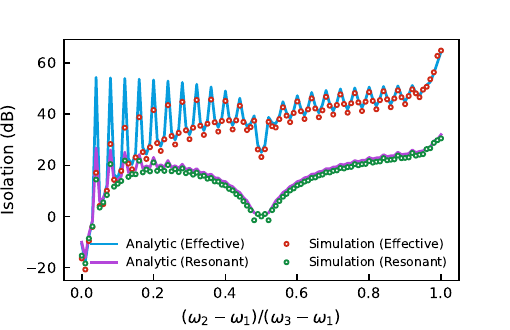}
\caption{
Isolation versus normalized intermediate frequency $(\omega_{2}-\omega_{1})/(\omega_{3}-\omega_{1})$. 
Blue line denote the analytic prediction for the effective model with separate gating windows, 
purple line show the analytic resonant model with overlapping couplings, 
red circles represent simulations based on the effective Hamiltonian, 
and green circles correspond to simulations of the full resonant Hamiltonian. 
Isolation remains nearly constant for the effective model but vanishes at the midpoint and rises toward the edges for the resonant case.}
  \label{fig:iso-vs-omega3}
\end{figure}

\paragraph{Intermediate mode frequency $\omega_{2}$.}
Fig.~\ref{fig:iso-vs-omega3} presents the isolation as a 
function of the intermediate frequency $\omega_{2}$, normalized 
by $(\omega_{3}-\omega_{1})$. 
Across the intermediate regime, specifically for normalized 
frequency ratios $0.05 < (\omega_2 - \omega_1)/(\omega_3 - \omega_1) 
< 0.95$, the effective sequential model predicts nearly constant 
isolation, as its dynamics are primarily governed by the gating 
period and duty cycle, and thus remain insensitive to the precise 
placement of $\omega_{2}$. 
In contrast, the resonant model exhibits a distinct minimum around 
the midpoint $\omega_{2}\!\approx\!(\omega_{1}+\omega_{3})/2$, where 
$\Omega_{1}=\omega_{2}-\omega_{1}$ and $\Omega_{2}=\omega_{3}-\omega_{2}$ 
coincide. At this point both couplings are simultaneously resonant, 
restoring approximate time-reversal symmetry and causing isolation 
to vanish. As $\omega_{2}$ shifts away from the center the two 
coupling frequencies separate, breaking this symmetry and yielding 
increased isolation.

At more extreme detunings two competing effects emerge. 
When $\Omega_{1}=\omega_{2}-\omega_{1}$ becomes comparable to $\Omega$, 
the Floquet harmonics mediating the $1\!\leftrightarrow\!2$ transition 
no longer uniquely select the forward pathway, so forward and reverse 
processes become simultaneously near-resonant and isolation decreases. 
Conversely, as $\Omega_{2}=\omega_{3}-\omega_{2}$ approaches zero the 
$2\!\leftrightarrow\!3$ coupling becomes effectively resonant during 
its gate window, strengthening the forward transfer without reopening 
the reverse pathway and increasing isolation. 
In both extreme limits the assumptions of well-separated coupling 
windows and weak spectral overlap no longer hold, so the simplified 
gating scheme may fail to reproduce the intended modulation pattern.

\begin{figure}[t]
  \centering
  \includegraphics[width=\linewidth]{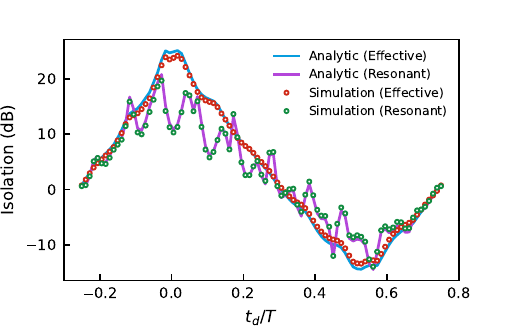}
\caption{
Isolation versus normalized delay between the two coupling gates for $D_{1} = D_{2} = 0.25$. 
The blue line shows the analytic prediction with separate gating windows, the purple line represents the resonant analytic prediction, red circles correspond to simulation using the effective Hamiltonian, and green circles indicate simulation with the resonant Hamiltonian. 
Maximum isolation occurs slightly before zero delay, where the forward path is maintained with minimal dwell time in the lossy intermediate mode, while larger delays or overlaps disrupt the timing, leading to oscillatory reduction and eventual reversal of directionality.}
  \label{fig:iso_delay}
\end{figure}

\paragraph{Relative delay between $D_{1}$ and $D_{2}$.}
Fig.~\ref{fig:iso_delay} shows the isolation as a function 
of the relative delay $t_d$ between the two sequential gating 
windows, defined as
\begin{equation}
g_{12}(t) = g_{12\left[\frac{t_d}{2},\,\tau_{1}+\frac{t_d}{2}\right)}(t), \qquad
g_{23}(t) = g_{23\left[\tau_{1}-\frac{t_d}{2},\,\tau_{1}-\frac{t_d}{2}+\tau_{2}\right)}.
\end{equation}
Positive values of $t_d$ introduce a gap between the two 
couplings, whereas negative values lead to partial overlap.
The maximum isolation occurs slightly before $t_d = 0$, where 
the two gates are sequential with minimal separation, allowing 
energy to flow along the forward path with less 
waiting in the lossy intermediate mode, while the reverse path 
is strongly attenuated.
As $|t_d|$ increases in either direction, the temporal 
sequencing required for directionality degrades.
For negative $t_d$, the windows progressively overlap until 
at $t_d = -\tau_1 = -\tau_2$ they coincide completely, 
restoring time-reversal symmetry and causing isolation to 
vanish, as seen in Fig.~\ref{fig:iso_delay}.
For positive $t_d$, the growing gap forces the excitation to 
dwell longer in the lossy intermediate mode, and as the gating 
order effectively reverses, the forward path is suppressed 
while energy from mode~3 is efficiently converted back to 
mode~1, leading to a reversal of the isolation direction.

\begin{figure}[t]
  \centering
  \includegraphics[width=\linewidth]{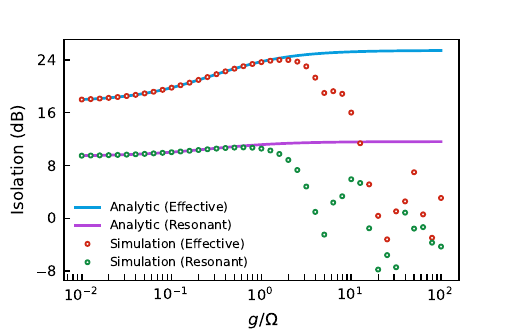}
\caption{Isolation versus normalized coupling strength $g/\Omega$, where $g = g_{12} = g_{23}$. 
Solid blue and purple lines show the analytical results for the effective and resonant models, respectively, while red circles and green squares represent the corresponding time-domain simulations. 
As $g$ increases, energy transfer between modes becomes more efficient and isolation rises until the perturbative limit of the model is reached. 
Beyond this point, higher-order interactions emerge and the analytic approximation loses accuracy, leading to gradual deviation and saturation in the simulated response.}
  \label{fig:iso-g}
\end{figure}

\paragraph{Coupling strength $g$.}
Fig.~\ref{fig:iso-g} shows the dependence of isolation on the normalized coupling strength $g/\Omega$. 
As the coupling strength increases, the energy exchange rate between neighboring modes accelerates, enabling more efficient transfer through the sequential pathway. 
In this weak-coupling regime, the isolation rises with $g$, and both analytic models accurately capture this trend. 
At stronger coupling, however, the assumptions underlying the perturbative treatment break down. 
Higher-order mixing between sidebands becomes significant, and the truncated series used in the analytic model no longer converges. 
In the time-domain simulations, these effects appear as a gradual deviation and eventual saturation of isolation. Beyond this limit, the system departs from the perturbative regime.

\section{Discussions of the approach}
The proposed framework provides an efficient and physically transparent method for predicting nonreciprocal frequency conversion within a controlled parameter regime. Its validity is restricted to cases where the series expansion in Eq.~(\ref{eq:Final_R}) remains convergent. When this expansion breaks down, such as in the presence of very large loss in the intermediate mode, extremely small modulation frequencies $\Omega$, or sufficiently strong coupling strengths $g$ as shown in Fig.~\ref{fig:iso-g}, an increasingly large number of higher-order terms contributes to the response. In this regime, a truncated analytical description becomes inaccurate and direct numerical simulations are required.

In addition, while sharper modulation pulses corresponding to smaller duty cycles $D$ generally enhance isolation by increasing temporal asymmetry between forward and reverse processes, they also move the effective coupling away from optimal pulse-area conditions. As a result, excessive pulse sharpening can increase insertion loss in the forward direction due to reduced conversion efficiency and enhanced dissipation. From an experimental perspective, achieving very small duty cycles further introduces challenges related to timing precision, waveform fidelity, and the preservation of frequency-selective coupling, which may limit the practical operating regime of the protocol.

\section{experimental implementation}
Our approach could be experimentally implemented on various platforms to enable nonreciprocal frequency conversion. 
Examples include integrated optical \cite{zhang2019electronically} and mechanical \cite{jun} devices, where resonant modes emulate atomic energy levels, as illustrated in Fig.~\ref{fig:schem-two-panels} (a). 
The transitions between different resonant modes could leverage electro-optic \cite{zhu2021integrated}, photoelastic \cite{eichenfield2009optomechanical}, or electro-acoustic effects \cite{shao2022electrical}.

Beyond this, our framework applies to quantum systems such as superconducting circuits \cite{krantz2019guide}, where tunable couplers allow for precise time-domain modulation of inter-qubit interactions. Another promising candidate is solid-state defects, such as nitrogen-vacancy centers in diamond \cite{doherty2013nitrogen}. In these systems, the transitions between spin sublevels form a naturally discrete three-mode system that can be coherently driven by sequenced microwave pulses \cite{jelezko2004observation}.

We recently demonstrated this nonreciprocal frequency conversion between gigahertz mechanical modes in lithium niobate phononic crystals ~\cite{jun}. The transitions between modes have been realized by electrical modulation, and a maximum non-reciprocity of 20 dB has been experimentally demonstrated. 

\section{Conclusion}
We analyzed nonreciprocal frequency conversion in a three-mode system with sequential time-gated couplings using harmonic-balance and Dyson--Born theory, validated by direct time-domain simulations.
Across all approaches, the underlying mechanism is clear: in the forward direction energy transfers sequentially through the system, while in the reverse direction the excitation is forced to dwell in the lossy intermediate mode~2, resulting in isolation.
This dwell-time asymmetry provides a simple physical picture that unifies the behavior observed across parameters.

We showed that isolation can be tuned and optimized by adjusting the modulation frequency $\Omega$, duty cycles $D_1$ and $D_2$, relative delay between the gates, and the intermediate mode loss $\kappa_2$. 
Smaller duty cycles and properly timed gates promote sequential energy transfer between modes, thereby strengthening directionality.
Moderate loss enhances isolation by penalizing the reverse path, while excessively large $\kappa$ suppresses amplitudes in both directions.
Similarly, appropriate choice of $\Omega$ balances dwell-time suppression and sideband strength, and fine control of delay governs whether the gates cooperate or compete in phase.

Together, these results provide a unified framework and clear design guidelines for realizing magnet-free isolators.
The approach is general and directly applicable to diverse platforms, offering a fast analytical route to explore new time-modulated nonreciprocal architectures.

\section*{Acknowledgements}
J.J. and L.S. were partially supported by the Air Force Office of Scientific Research (AFOSR) under Grant Number W911NF-23-1-0235. The views and conclusions contained in this document are those of the authors and do not necessarily reflect the position or the policy of the United States Government. No official endorsement should be inferred. Approved for public release.

\section*{Data Availability}
Source codes and data that supports the findings of this study are available at DOI: \url{10.6084/m9.figshare.31254754}.

For review, please visit the following link to access data and source codes:
\url{https://figshare.com/s/ad5282c7c6c49aee916c}

\bibliography{references}

\appendix

\section{Harmonic-Balance Formulation and Born Expansion}
\label{Apx:HB}

This appendix details the harmonic-balance formalism used to solve the time-periodic equations of motion and provides the derivation of the Dyson-Born series expansion utilized in the main text.

\subsection{Fourier representation and recurrence relation}
We begin by expanding the time-periodic interaction Hamiltonian $V(t)$ into its Fourier components and defining the diagonal bare resolvent (Green's function) for the uncoupled system:
\[
V(t)=\sum_{m\in\mathbb Z}V^{(m)}\,e^{-im\Omega t},\qquad
\bm G_0(\omega):=(\omega\,\bm I-H_0)^{-1}.
\]
Projecting the time-domain equation of motion onto the $n$-th Floquet sideband yields the fundamental harmonic-balance (HB) recursion relation:
\begin{equation}
\bm a^{(n)}=\bm G_0(\omega_p{+}n\Omega)\Big(\sum_{m}V^{(m)}\,\bm a^{(n-m)}+F\,\delta_{n0}\,\bm u\Big).
\label{eq:HB_recursion_apx}
\end{equation}

\subsection{Lift to Floquet space}
To solve this recursion, we lift the system into a global Floquet space. We construct a supervector $\bm A$ containing amplitudes across all sidebands, $\bm A:=(\ldots,\bm a^{(-1)},\bm a^{(0)},\bm a^{(+1)},\ldots)^\top$, and introduce the corresponding block-diagonal propagator and Toeplitz interaction matrix:
\begin{gather*}
\big(\bm{\mathcal{G}}_0\big)_{nn'}=\delta_{nn'}\,\bm G_0(\omega_p{+}n\Omega),\quad
\big(\bm{\mathcal{V}}\big)_{nn'}=V^{(n-n')},\\
\big(\bm{\mathcal{B}}\big)_{n}=F\,\delta_{n0}\,\bm u.
\end{gather*}
In this representation, the infinite set of coupled equations [Eq.~(\ref{eq:HB_recursion_apx})] condenses into a single linear algebraic equation:
\[
\bm A=\bm{\mathcal{G}}_0\big(\bm{\mathcal{V}}\,\bm A+\bm{\mathcal{B}}\big).
\]

\subsection{Lippmann-Schwinger solution}
Rearranging the linear equation isolates the state vector $\bm{A}$, yielding the formal Lippmann-Schwinger solution:
\[
(\bm{\mathcal{I}}-\bm{\mathcal{G}}_0\bm{\mathcal{V}})\bm A=\bm{\mathcal{G}}_0\bm{\mathcal{B}}
\quad\Longrightarrow\quad
\bm A=(\bm{\mathcal{I}}-\bm{\mathcal{G}}_0\bm{\mathcal{V}})^{-1}\,\bm{\mathcal{G}}_0\,\bm{\mathcal{B}}.
\]

\subsection{Perturbative expansion}
Under the condition of weak coupling or large detuning---specifically when the spectral radius $\rho(\bm{\mathcal{G}}_0\bm{\mathcal{V}})<1$---the inverse operator can be expanded as a Neumann series:
\[
(\bm{\mathcal{I}}-\bm{\mathcal{X}})^{-1}=\sum_{k=0}^{\infty}\bm{\mathcal{X}}^{k}\quad(\text{where }\bm{\mathcal{X}}:=\bm{\mathcal{G}}_0\bm{\mathcal{V}}).
\]
Substituting this back into the solution for $\bm{A}$ generates the Dyson series~\cite{Fetter1971}:
\[
\bm A=\sum_{k=0}^{\infty}(\bm{\mathcal{G}}_0\bm{\mathcal{V}})^{k}\,\bm{\mathcal{G}}_0\,\bm{\mathcal{B}}
=\bm{\mathcal{G}}_0\bm{\mathcal{B}}+\bm{\mathcal{G}}_0\bm{\mathcal{V}}\bm{\mathcal{G}}_0\bm{\mathcal{B}}
+\bm{\mathcal{G}}_0\bm{\mathcal{V}}\bm{\mathcal{G}}_0\bm{\mathcal{V}}\bm{\mathcal{G}}_0\bm{\mathcal{B}}+\cdots.
\]

\subsection{Extraction of harmonic components}
Projecting the block-vector equation back onto the specific $n$-th harmonic subspace recovers the iterative contributions to the sideband amplitude $\bm a^{(n)}$:
\begin{widetext}
\[
\begin{aligned}
\bm a^{(n)}_{(0)} &= \bm G_0(\omega_p{+}n\Omega)\,F\,\delta_{n0}\,\bm u, \\[6pt]
\bm a^{(n)}_{(1)} &= \bm G_0(\omega_p{+}n\Omega)\,V^{(n)}\,\bm G_0(\omega_p)\,F\bm u, \\[6pt]
\bm a^{(n)}_{(2)} &= \bm G_0(\omega_p{+}n\Omega)\!\sum_{m}V^{(m)}\,\bm G_0(\omega_p{+}(n{-}m)\Omega)\,V^{(n-m)}\,\bm G_0(\omega_p)\,F\bm u.
\end{aligned}
\]
\end{widetext}
The term $\bm a^{(n)}_{(2)}$ represents the lowest-order non-vanishing contribution to the sequential two-hop transport discussed in Sec.~\ref{sec:theory}.


\section{Time-Domain Simulation of Floquet Sidebands}
\label{Apx:sim}

This appendix details the numerical procedure used to simulate the dynamics of the driven three-mode system and to extract the steady-state Floquet sideband amplitudes via Fourier analysis.

\subsection{Model and Drive}
The system is characterized by bare modal frequencies $(\omega_1,\omega_2,\omega_3)$ and decay rates $(\kappa_1,\kappa_2,\kappa_3)$. The pump frequency $\omega_p$ is chosen to match the input mode of the directional process:
\[
\omega_p =
\begin{cases}
\omega_1, & \text{forward case ($1\to 3$ conversion)},\\
\omega_3, & \text{reverse case ($3\to 1$ conversion)}.
\end{cases}
\]

\paragraph{Transformation to the rotating frame.}
In the laboratory frame, the modal amplitudes oscillate rapidly at the carrier frequency. To eliminate this fast carrier oscillation, we transform to a frame rotating at $\omega_p$:
\[
a_j(t) = b_j(t)\,e^{-i\omega_p t}, \qquad j\in\{1,2,3\},
\]
where $b_j(t)$ evolve on timescales determined by the detunings and coupling rates. The laboratory-frame equation of motion,
\[
i\,\dot{\bm a}(t) = H(t)\,\bm a(t) + iF\,e^{-i\omega_p t}\,\bm u,
\]
transforms into
\[
i\,\dot{\bm b}(t) = H'(t)\,\bm b(t) + iF\,\bm u,
\]
with the effective Hamiltonian
\begin{equation}
H'(t) =
\begin{bmatrix}
\Delta_1 & -V_{12}(t) & 0\\
-V_{12}(t) & \Delta_2 & -V_{23}(t)\\
0 & -V_{23}(t) & \Delta_3
\end{bmatrix},
\end{equation}
where the diagonal elements $\Delta_j = (\omega_j - \omega_p) - i\kappa_j/2$ account for both detuning and dissipation; the imaginary part of each diagonal entry renders $H'(t)$ non-Hermitian. In this frame, the external drive appears as a static source term $iF\bm{u}$.

\paragraph{Sequential time-gated couplings.}
The time-dependent couplings $V_{12}(t)$ and $V_{23}(t)$ are activated in sequential, non-overlapping windows of duration $\tau_1 = D_1 T$ and $\tau_2 = D_2 T$, respectively, within the modulation period $T=2\pi/\Omega$:
\begin{align}
V_{12}(t) &= g_{12}(t)\,\cos(p_1\Omega t),\\
V_{23}(t) &= g_{23}(t)\,\cos(p_2\Omega t).
\end{align}
The envelope functions are defined as rectangular pulses:
\[
g_{12}(t) = \begin{cases} g_{12} & 0 \le t < \tau_1 \\ 0 & \text{elsewhere} \end{cases},
\quad
g_{23}(t) = \begin{cases} g_{23} & \tau_1 \le t < \tau_1 + \tau_2 \\ 0 & \text{elsewhere} \end{cases}.
\]
The integers $p_1$ and $p_2$ are selected such that $p_1\Omega \approx \omega_2-\omega_1$ and $p_2\Omega \approx \omega_3-\omega_2$, ensuring near-resonant coupling.

\subsection{Numerical Integration}
We integrate the differential equation
\[
\dot{\bm b}(t) = -i\,H'(t)\,\bm b(t) + F\,\bm u
\]
using a fixed-step fourth-order Runge-Kutta (RK4) scheme. To ensure numerical stability and accuracy, the time step $dt$ is chosen to resolve the fastest dynamical timescale:
\[
dt \le \frac{2\pi}{\mathrm{PPW} \cdot \omega_{\max}},
\]
where $\omega_{\max} = \max\{p_1\Omega,\, p_2\Omega,\, |\omega_3-\omega_1|\}$ and $\mathrm{PPW}$ (points per wavelength) is an oversampling factor (typically $\ge 20$). Transients are eliminated by propagating the system for several decay lifetimes ($t \gg 1/\min(\kappa_j)$), and steady state is verified by comparing $\bm b(t)$ across successive modulation periods.

\subsection{Floquet Amplitude Extraction}
Once steady state is established, the trajectory $\bm b(t)$ is recorded over a single period $T$. The Floquet sideband amplitudes are extracted via the Fourier integral:
\begin{equation}
a_j^{(n)} = \frac{1}{T}\int_{t_0}^{t_0+T} b_j(t)\,e^{in\Omega t}\,dt, \qquad j\in\{1,2,3\},
\end{equation}
where we consider harmonics $n \in [-N_{\mathrm{harm}}, \ldots, N_{\mathrm{harm}}]$. These amplitudes correspond to the spectral components at absolute frequencies $\omega_n = \omega_p + n\Omega$. In the spectral plots, we define the frequency axis relative to the input mode, plotting vs.\ $(\omega - \omega_1)/\omega_1$.

\end{document}